\documentclass[twocolumn,aps,prx,superscriptaddress,floatfix]{revtex4-2}
\usepackage[dvipsnames]{xcolor}
\usepackage{comment}
\usepackage[T1]{fontenc}
\usepackage[utf8]{inputenc}
\usepackage[english]{babel}
\usepackage{multibib}
\usepackage{color}
\usepackage{float}
\usepackage{amsmath}
\usepackage{amstext}
\usepackage{amssymb,bm}
\usepackage{graphicx}
\usepackage{bbold}
\usepackage{hyperref}
\usepackage{array}[=2016-10-06]

\newcommand{\be}{\begin{equation}}
\newcommand{\ee}{\end{equation}}

\def\f0{f^0_\mathbf{k}}

\def\*#1{\mathbf{#1}}

\renewcommand{\vec}{\mathbf} 
\newcommand{\eps}{\varepsilon}
\def\Re{\text{Re}}
\def\Im{\text{Im}}

\newcommand{\db}[1]{\overline{{\bf #1}}} 

\usepackage{soul}

\setul{0.5ex}{0.3ex}
\definecolor{Blue}{rgb}{0,0.0,1}
\setulcolor{Blue}
\usepackage{babel}

\begin{document} 
\author{Sylvain Lanneb\`ere}
\thanks{These authors contributed equally to this work.}
\affiliation{Instituto de Telecomunicações and Department of Electrical and Computer Engineering, University of Coimbra, 3030-290 Coimbra, Portugal}

\author{Tatiana G. Rappoport}
\thanks{These authors contributed equally to this work.}
\affiliation{Centro Brasileiro de Pesquisas Físicas (CBPF), Rua Dr Xavier Sigaud 150, Urca, 22290-180, Rio de Janeiro-RJ, Brazil}
\affiliation{International Iberian Nanotechnology Laboratory (INL), Av. Mestre José Veiga, 4715-330 Braga, Portugal}

\author{Tiago A. Morgado} 
\affiliation{Instituto de Telecomunicações and Department of Electrical Engineering, University of Coimbra, 3030-290 Coimbra, Portugal} 

\author{Ivo Souza} 
\affiliation{Centro de Física de Materiales, Universidad del País Vasco, 20018 San Sebastián, Spain}
\affiliation{Ikerbasque Foundation, 48013 Bilbao, Spain}
\author{M\'ario G. Silveirinha}
\affiliation{University of Lisbon and Instituto de Telecomunicações, Avenida Rovisco Pais 1, Lisboa, 1049-001 Portugal}	

\title{Symmetry Analysis of the Non-Hermitian Electro-Optic Effect in Crystals
}
\begin{abstract}

Here, we investigate how crystal symmetry tailors the non-Hermitian electro-optic effect arising from the Berry curvature dipole. Specifically, we demonstrate the critical influence of the material's point group symmetry and external electric biases in shaping this effect, leading to current-induced optical gain and non-reciprocal optical responses. Through a symmetry-based analysis of the crystallographic point groups, we identify how different symmetries affect the electro-optic response, enabling the engineering of polarization-dependent optical gain without the need for gyrotropic effects. In particular, we demonstrate that the non-Hermitian electro-optic response in a broad class of crystals is characterized by linear dichroic gain. In this type of response, the eigenpolarizations that activate the gain or dissipation are linearly polarized. Depending on the point group symmetry, it is possible to achieve gain (or dissipation) for all eigenpolarizations or to observe polarization-dependent gain and dissipation.
Weyl semimetals emerge as promising candidates for realizing significant non-Hermitian electro-optic effects and linear dichroic gain. We further examine practical applications by studying the reflectance of biased materials in setups involving mirrors, demonstrating how optical gain and attenuation can be controlled via symmetry and bias configurations. 
 \end{abstract}

\maketitle
\section{Introduction} 

Non-linear optical and transport phenomena are hallmark features of materials with broken symmetries, leading to unique effects such as second-harmonic generation (SHG) \cite{Sodemann2015,Xu2018,Ye2023}, the non-linear Hall effect \cite{Ma2018,Kang2019,NonlinearHall2021} and rectification \cite{Zhang2021,suarez-rodriguez_odd_2024}. These phenomena are profoundly influenced by the material’s underlying symmetries. As a result, non-linear responses, including optical and transport effects, provide a fertile platform for exploring how material's quantum geometry and symmetry shape emergent quantum phenomena.

At the core of several of these non-linear effects lies the Berry curvature: a geometric property of electronic band structures that governs charge and spin transport. The (Abelian) Berry curvature is different from zero only in systems where either time-reversal symmetry (TRS) or inversion symmetry is broken \cite{Xiao2010,vanderbilt_berry_2018}. The Berry curvature dipole (BD), which represents the dipole moment of the Berry curvature in momentum space \cite{Deyo2009,Sodemann2015,Xu2018,Zhang2021}, plays a pivotal role in non-linear transport phenomena such as the non-linear Hall effect and other second-order responses. The BD enables the generation of a transverse current, perpendicular to the applied electric field, without requiring an external magnetic field or internal magnetic order, as demonstrated in several low-symmetry materials \cite{Ma2018,you_berry_2018,Kang2019,NonlinearHall2021}. Systems with a finite BD also exhibit the kinetic magnetoelectric effect, where an electric current generates a net magnetization in the material \cite{Shalygin2012,Furukawa2017,son_strain_2019,Calavalle2022}.

Several studies have further linked the BD to optical effects, including SHG, the kinetic Faraday effect, where the rotatory power is proportional to the electric current and reverses sign with the applied electric field \cite{Vorobev1979,Tsirkin2018, Konig2019,fu_berry_2023}, and the circular photogalvanic effect, where the photocurrent reverses sign with the helicity of light \cite{Ivchenko1978,Asnin1978,Deyo2009,Xu2018}. These effects are particularly prominent in  materials with small band gaps, where the Berry curvature gets resonantly enhanced and significantly influences interband transitions. Crucially, many second-order optical responses have been shown to directly correlate with the BD, establishing them as effective probes for investigating a material’s electronic structure and symmetry \cite{Xu2018,bhalla_resonant_2022}. 

While the nonlinear Hall effect primarily describes how the anomalous velocity rectifies a time-varying field to generate a DC current, with the response characterized by the BD, more recent studies have explored its role in the electro-optic response. In particular, it has been shown that electro-optic effects can give rise to both nonreciprocity and optical gain, where the amplification of an optical signal in a biased medium depends on both the light’s polarization and its direction of travel \cite{mosfet,Rappoport2023,Shi2022, Morgado2024,Hakimi2024,desousa2024,lannebere_chiral_2025,Ma2025,roldan_optical_2025,eslami_berry_2026}. This effect is tied to the Berry curvature of the system and has been shown to produce non-Hermitian (NH) electro-optic responses, leading to current-induced optical gain in both two- and three-dimensional materials \cite{Rappoport2023, Morgado2024}. This phenomenon can also be explored in the context of lasing, demonstrating that NH effects can lead to polarization-dependent gain in biased systems \cite{Hakimi2024,Morgado2024,lannebere_chiral_2025}. These works highlight the role of Berry curvature dipoles in shaping electro-optic effects, particularly in materials with low symmetry.

Building on our previous studies of the non-Hermitian electro-optic (NHEO) effect \cite{mosfet,Rappoport2023,Morgado2024,Hakimi2024,lannebere_chiral_2025}, here we investigate how crystalline point group symmetries and bias configurations shape the non-conservative response. 
In three dimensions, the BD is a (pseudo)tensor that determines the eigenpolarizations associated with both gain and dissipative responses in the NHEO effect. For instance, in materials belonging to point group 32, such as tellurium, we previously demonstrated that the eigenpolarizations responsible for gain and dissipation are circularly polarized and exhibit opposite handednesses \cite{Morgado2024}. In this study, we show that by precisely engineering the Berry curvature dipole tensor via the intrinsic crystalline point group symmetry, it is possible to control the polarization type—linear, circular, or elliptical—as well as the corresponding handedness that drives these responses.
Moreover, we develop a comprehensive ``symmetry roadmap'' that systematically links the crystal symmetries of various materials to their specific non-Hermitian and nonreciprocal responses under static electric bias. Notably, we identify a broad range of point groups that enable linear dichroic gain, where the gain or dissipation is governed by linearly polarized fields.
We remind the reader that nonreciprocity refers to responses that enable asymmetric propagation between two points, whereas non-Hermitian interactions are associated with dissipative or gain-related processes.

To perform the symmetry analysis, we investigate the allowed components of the BD tensor with the tools implemented in the Bilbao Crystallographic Server (BCS) \cite{Aroyo2006,Aroyo2006b}. 
We explore various symmetries and examine different electric bias configurations.   
Notably, we demonstrate that the NHEO effect can be engineered to produce optical gain for all polarizations of the optical field (in the plane perpendicular to the bias), which can be switched to optical dissipation simply by reversing the orientation of the static electric bias. Our analysis also reveals that gain can occur in the absence of any gyrotropic effect (optical rotation).

Finally, we illustrate the practical implications of the NHEO effect by examining the reflectance of materials in setups involving mirrors. This allows us to evaluate how the incoming wave polarization can trigger either dissipative or gain responses in systems with different symmetries, providing valuable insights into the potential applications of these materials in optical devices and electromagnetic systems.

\section{Electro-optic permittivity}
The linearized optical response of a generic low-symmetry three-dimensional conductor under a static electric bias $\*E_0$ can be found using the semiclassical Boltzmann transport theory   \cite{Rappoport2023, Morgado2024}.  The linear electro-optic response is determined by the BD tensor $\db{D}$, 
\begin{equation}
D_{ij}=-\frac{1}{(2\pi)^3}\int\frac{\partial f^0_\*k}{\partial k_i}\Omega_{\*k, j} d^3{k}=\frac{1}{(2\pi)^3}\int f^0_\*k\frac{\partial{\Omega_{\*k, j}}}{\partial k_i} d^3{k},
\end{equation}
where $ f^0_\*k$ is the Fermi-Dirac distribution and $\bm{\Omega}_\*k$ is the Berry curvature (with implied band summations). The BD in 3D materials is dimensionless and traceless, so that $\sum_i D_{ii}
=0$.   
 The electro-optic conductivity can be expressed as a sum of two terms, \cite{Morgado2024}:
\begin{align}\label{E:conductivity_EO}
\db{\boldsymbol{\sigma}}_\text{EO}(\omega)=\frac{ e^3\tau}{\hbar^2  } \left(- \vec{E}_{0}\cdot   \db{D} \times \vec{1}_{3\times3} +  \frac{1}{1-i\omega\tau}\vec{E}_{0}   \times \db{D}^T \right)
\end{align}
where $\tau$ is the scattering relaxation time, and $\db{D}^T$ denotes the transpose of the BD tensor. The first  term is associated with conservative light-matter interactions, while the second term    is associated with non-conservative interactions.

It is convenient to express the optical response in terms of an EO susceptibility (permittivity):
\begin{align}  \label{E:permittivity_EO}
\db{\eps}_\text{EO}(\omega)=\frac{\db{\boldsymbol{\sigma}}_\text{EO}}{-i\eps_0\omega}&=\db{\eps}_\text{EO}'+i\db{\eps}_\text{EO}''
\end{align}
where $\db{\eps}_\text{EO}'= \frac{\db{\eps}_\text{EO} + \db{\eps}_\text{EO}^\dag}{2}$ is the Hermitian part of the susceptibility and $i\db{\eps}_\text{EO}''= \frac{\db{\eps}_\text{EO} - \db{\eps}_\text{EO}^\dag}{2}$ is the anti-Hermitian part. Here, $\dag$  represents  the Hermitian-conjugate operation. Both  $\db{\eps}_\text{EO}'$ and  $\db{\eps}_\text{EO}''$ are Hermitian tensors with real-valued eigenvalues.  
The Hermitian component, $\db{\eps}_\text{EO}'$, contributes to the conservative part of the  material response, dictating the wave dispersion, while the anti-Hermitian part, $i\db{\eps}_\text{EO}''$, controls the power exchange between the wave and the medium.  

It is useful to decompose the two tensors into real (subscript R) and imaginary (subscript I) parts. For both $\db{\eps}_\text{EO}'$ and $\db{\eps}_\text{EO}''$, the real part is a symmetric tensor, while the imaginary part is an antisymmetric tensor. As shown in Appendix \ref{sec:appendix_electro_optic}, they can be written explicitly in terms of the (vectorial) frequency $\boldsymbol{\omega}_0=\frac{ e^3\tau}{\eps_0 \hbar^2  } \vec{E}_{0}$ as follows,
\begin{subequations} \label{eqn:epsEOdecomp}
\begin{align} 
\db{\eps}_\text{EO,R}'=& \frac{-\tau}{2 \left[1+(\omega\tau)^2\right]} |\boldsymbol{\omega}_0|\db{D}^{\rm S}_ {\boldsymbol{\omega}_0}, \label{E:epsPrimeReal}\\
\db{\eps}_\text{EO,I}'=&
   -\left(\frac{1}{\omega} +\frac{1}{2 \omega \left[1+(\omega\tau)^2\right]} \right) \left(\db{D}^T \cdot \boldsymbol{\omega}_0  
\right) \times \boldsymbol{1}_{3\times3},  \label{E:epsPrimeImag}\\
\db{\eps}_\text{EO,R}''=&   \frac{1}{2\omega\left[1+(\omega\tau)^2\right]}|\boldsymbol{\omega}_0| \db{D}^{\rm S}_ {\boldsymbol{\omega}_0},\label{E:epsSecondReal}\\
\db{\eps}_\text{EO,I}''=&\frac{-\tau }{2\left[1+(\omega\tau)^2\right]} \left(\db{D}^T \cdot \boldsymbol{\omega}_0  
\right) \times \boldsymbol{1}_{3\times3}. \label{E:epsSecondImag}
\end{align}
\end{subequations}
We defined $ \db{D}^{\rm S}_ {\boldsymbol{\omega}_0}$ as the real symmetric tensor $ \db{D}^{\rm S}_ {\boldsymbol{\omega}_0} \equiv  (\boldsymbol{\omega}_0   \times \db{D}^T -\db{D} \times \boldsymbol{\omega}_0)/|\boldsymbol{\omega}_0| $.
It is interesting to note that $\db{\eps}_\text{EO,R}'$ and $\db{\eps}_\text{EO,R}''$ have the same structure but different frequency dependence. Similarly, $\db{\eps}_\text{EO,I}'$ and $\db{\eps}_\text{EO,I}''$ also share the same structure and a different frequency variation.

\section{Chiral-gain and Linear-dichroic gain}
\label{Sec:EOgain}

In table \ref{table}, we summarize how the different tensors correlate to different electromagnetic properties of the material. 
The component $\db{\eps}_\text{EO,R}'$ represents a common conservative (Hermitian) reciprocal response. It typically induces or enhances birefringence, causing the refractive index experienced by a wave to depend on its direction of propagation and polarization. On the other hand, $i \db{\eps}_\text{EO,I}'$ represents a conservative nonreciprocal ($\db{\eps} \neq \db{\eps}^T$) gyrotropic response, analogous to the response of a magnetized plasma. The equivalent bias magnetic field is oriented parallel to the direction $-\db{D}^T \cdot \boldsymbol{\omega}_0$ \cite{Morgado2024}.
This gyrotropic response also implies time-reversal symmetry breaking, as $\db{\eps} \neq \db{\eps}^*$. 
\begin{table}[h!]
\begin{tabular}{ |c|c|c| } 
 \hline
  & {\bf Reciprocal} & {\bf Non-Reciprocal} \\
 \hline
 {\bf Conservative} & $\db{\eps}_\text{EO,R}'$ (Birefringence) & $i \db{\eps}_\text{EO,I}'$ (Gyrotropic) \\ 
 \hline
{\bf  Non-conservative} & $\db{\eps}_\text{EO,R}''$ (Linear gain) & $i\db{\eps}_\text{EO,I}''$ (Chiral gain)\\ 
 \hline
\end{tabular}
\caption{Classification of the real and imaginary components of $\db{\eps}_\text{EO}'$ and $\db{\eps}_\text{EO}''$ based on their reciprocal and non-reciprocal, conservative and non-conservative properties. The term chiral gain refers to circular-dichroic gain.}
\label{table}
\end{table}

Conversely, $\db{\eps}_\text{EO}''$ describes a non-conservative (non-Hermitian) response responsible for optical gain or loss. As discussed in our previous works \cite{ Morgado2024,Rappoport2023} the time-averaged net power exchanged between an electromagnetic wave and a material per unit of volume is
\begin{align}
\label{E:powerdip}
p_{\rm d} = \frac{\omega \eps_0}{2} \vec{E}_\omega^\ast \cdot \db{\eps}'' \cdot \vec{E}_\omega. 
\end{align}
where $i\db{\eps}''$ is the anti-Hermitian part of the full permittivity tensor $\db{\eps}$.
Here, it is implicit that the fields have a time-harmonic variation $\vec{E}(t) = \frac{1}{2}\left(\vec{E}_\omega e^{-i\omega t} + \vec{E}_\omega^\ast e^{i\omega t}\right)$. 
Thus, the quadratic form $\vec{E}_\omega^\ast \cdot \db{\eps}'' \cdot \vec{E}_\omega$ governs the direction of the irreversible energy flow. For a positive $\vec{E}_\omega^\ast \cdot \db{\eps}'' \cdot \vec{E}_\omega$ the energy of the wave is irreversibly dissipated in the material in the form of heat. Conversely, for a negative $\vec{E}_\omega^\ast \cdot \db{\eps}'' \cdot \vec{E}_\omega$ the medium supplies energy to the wave, corresponding to optical gain. The tensor $\db{\eps}_\text{EO}''$ describes the non-conservative interactions that arise from the electro-optic effect. It combines additively with the standard linear (Drude) response contribution of the material \cite{ Morgado2024,Rappoport2023}, which is strictly dissipative.

As previously noted, $\db{\eps}_\text{EO}''$ is an Hermitian tensor and thereby has a spectral decomposition of the type 
\begin{align}    
\db{\eps}_\text{EO}''
 = \sum\limits_{i = 1,2,3} {\lambda _i {\bf{u}}_i  \otimes {\bf{u}}_i^* },
\end{align}
where $\lambda _i$ are the real-valued eigenvalues, ${\bf{u}}_i$ are the  eigenvectors normalized as $
{\bf{u}}_i  \cdot {\bf{u}}_j^*  = \delta _{ij}$ and $\otimes$ represents the tensor product of two vectors. The contribution of the electro-optic response to the dissipated power can be expressed in terms of the eigenvalues as follows:
\begin{align}
p_{\rm d,EO} = \frac{\omega \eps_0}{2}
\sum\limits_{i = 1,2,3} {\lambda _i \left| {{\bf{E}}_\omega   \cdot {\bf{u}}_i^* } \right|^2 } 
. 
\end{align}
The sign of $p_{\rm d,EO}$ is determined by the sign of the eigenvalues $\lambda_i$. 
Specifically, when $\lambda_i < 0$ the corresponding eigenpolarization ${\bf{E}}_\omega   \sim {\bf{u}}_i$ activates gain in the material, whereas when $\lambda_i > 0$ the eigenpolarization activates dissipation.

From Eq. \eqref{eqn:epsEOdecomp}, we observe that
\begin{align}
\boldsymbol{\omega}_0 \cdot \db{\eps}_\text{EO}'' \cdot \boldsymbol{\omega}_0 = 0.
\end{align}
This indicates that $\db{\eps}_\text{EO}''$ cannot be a positive definite or negative definite tensor, meaning its eigenvalues cannot all share the same sign. Consequently, the light-matter interactions arising from the electro-optic effect are typically indefinite, meaning that the electro-optic contribution to the power dissipated in the material is governed by an indefinite quadratic form. Thus, the non-Hermitian response can be either active or dissipative, depending on the eigenpolarization.

In this work, we focus on material systems with a symmetry axis and a bias field aligned with that axis. For such systems, $\boldsymbol{\omega}_0$ is itself an eigenvector (${\bf{u}}_3 \sim \boldsymbol{\omega}_0$) and is necessarily associated with a trivial eigenvalue ($\lambda_3 = 0$). Therefore, in this class of platforms, the NHEO response is governed by two orthogonal eigenpolarizations, ${\bf{u}}_1$ and ${\bf{u}}_2$, which lie in the plane perpendicular to the system's symmetry axis.

Let us now focus on the contribution of the term $i\db{\eps}_\text{EO,I}''$ to $p_{\rm d,EO}$. Following Ref. \onlinecite{lannebere_chiral_2025}, it can be written explicitly as:
\begin{align}
p_{\rm d,EO}^{\rm I} = -\frac{\omega \eps_0}{2}
\boldsymbol{\sigma } \cdot \boldsymbol{\Omega}_\omega  \left| {{\bf{E}}_\omega  } \right|^2, 
\end{align}
where $
\boldsymbol{\sigma } = i\left( {{\bf{E}}_\omega   \times {\bf{E}}_\omega ^* } \right)/\left| {{\bf{E}}_\omega  } \right|^2 
$ is the spin angular momentum of the wave, whereas 
\begin{align}
\boldsymbol{\Omega}_\omega = \frac{\tau }{2\left[1+(\omega\tau)^2\right]} \db{D}^T \cdot \boldsymbol{\omega}_0
\end{align}
is by definition the chiral-gain vector. Thus, the light-matter interactions can be either dissipative ($p_{\rm d,EO}^{\rm I}>0$) or active ($p_{\rm d,EO}^{\rm I}<0$), depending on the relative alignment between the spin angular momentum of the wave and the chiral-gain vector. The optimal polarization to unlock the gain is a circular polarization in the plane perpendicular to the chiral-gain vector, whereas the optimal polarization to unlock the dissipative properties of the medium is a circular polarization in the same plane but with opposite handedness \cite{lannebere_chiral_2025}. Accordingly, the two eigenvectors of $i\db{\eps}_\text{EO,I}''$ are two circular polarizations in the plane perpendicular to $\boldsymbol{\Omega}_\omega$. As shown in table \ref{table},  $i\db{\eps}_\text{EO,I}''$ also contributes to a nonreciprocal electromagnetic response. Note also that as the gyrotropic and chiral-gain terms have the same mathematical structure, they typically coexist within the same material.

Next, we analyze the contribution of the term $\db{\eps}_\text{EO,R}''$ to $p_{\rm d}$. Since $\db{\eps}_\text{EO,R}''$ is real-valued, its eigenvectors ${\bf{u}}_i$ are necessarily real-valued as well. Unlike the chiral-gain case, where non-conservative effects are associated with circular polarizations, the non-conservative effects in this scenario are governed by linear polarizations with trivial spin angular momentum. We refer to this type of gain as ``linear dichroic gain'' to emphasize that the associated eigenpolarizations are linearly polarized and that the non-conservative effects depend on polarization. As shown in Table \ref{table}, the linear dichroic gain is associated with a reciprocal electromagnetic response.

Later in the article, we present examples where the linear dichroic gain is either indefinite—meaning the gain is activated by a specific linear polarization while dissipation occurs for an orthogonal linear polarization—or semi-definite, where the non-conservative effects (when nontrivial) are exclusively dissipative or active.

In the general case, where $\db{\eps}_\text{EO}''$ has both real and imaginary components, the eigenpolarizations that activate gain and dissipation are typically elliptically polarized. Furthermore, the frequency dependence of the real and imaginary parts of $\db{\eps}_\text{EO}''$ plays a critical role in determining optical gain. For both $\db{\eps}_\text{EO,R}''$ and $\db{\eps}_\text{EO,I}''$, the non-conservative effects become increasingly pronounced as $\omega \to 0$. In this low-frequency limit, the real (symmetric) part, $\db{\eps}_\text{EO,R}''$, dominates due to its $1/\omega$ dependence on frequency.
As a result, at lower frequencies, the non-conservative physics is predominantly influenced by linear dichroic gain. However, at higher frequencies, the scaling behavior changes: the linear dichroic gain decreases as $1/\omega^3$, while the chiral gain decreases more slowly as $1/\omega^2$. This indicates that at higher frequencies, chiral gain becomes the dominant contributor to the optical gain.

In the next section, we will examine the properties of $\db{\eps}_\text{EO}''$ across various point-group symmetries of non-centrosymmetric crystals. We will show that the structure of the BD tensor is intrinsically determined by the material's symmetry. As a result, the symmetry of the material plays a crucial role in shaping the non-conservative light-matter interactions arising from the electro-optic effect.

\section{Symmetry Analysis of the Non-Hermitian Response}
\begin{figure}
	
		\includegraphics[width=0.98\linewidth]{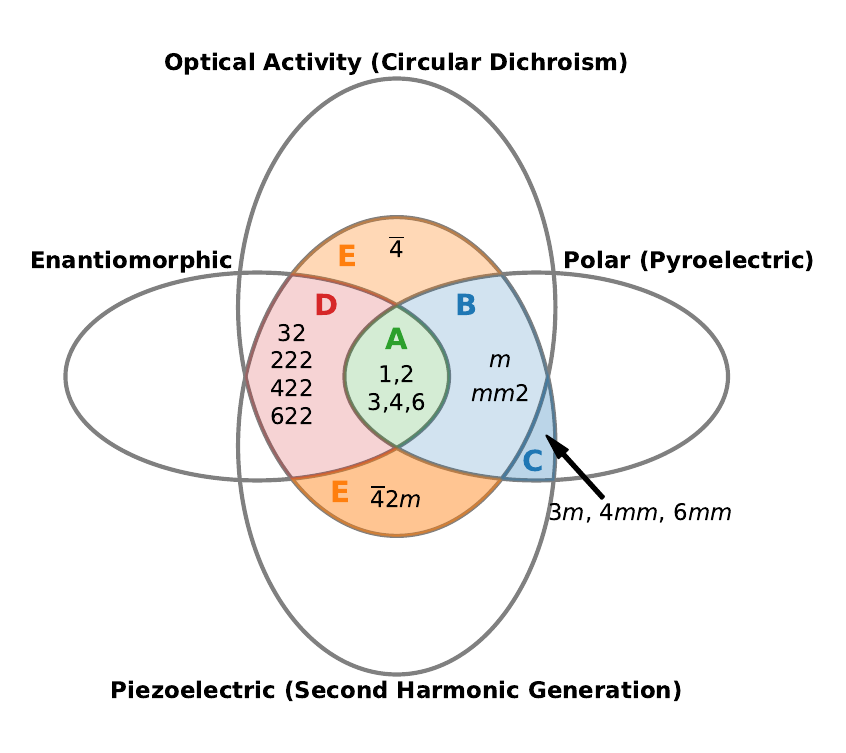}\caption{Classification diagram based on the work of Halasyamani and Poeppelmeier \cite{Halasyamani1998}, dividing non-centrosymmetric materials into seven categories according to their symmetry-dependent properties. Only the five categories that can have a non-zero BD are shown in the diagram. The diagram classifies materials by separating them into polar and nonpolar crystal sets, and also organizing them based on properties such as optical activity, circular dichroism, and piezoelectricity. The diagram also serves to categorize the BD tensor of various materials.}
\label{fig1}
\end{figure}
 As explained in Ref.\cite{Sodemann2015}, in three dimensions the BD $\db{D}$ transforms as a rank-two pseudotensor  and thus necessarily vanishes in the 11 crystallographic point groups (out of 32) that are centrosymmetric. 
Beyond inversion, additional symmetries further constrain the allowed components of $\db{D}$. 

By decomposing $\db{D}$ into symmetric and antisymmetric parts, one can show that the antisymmetric part of $\db{D}$ is nonzero only in polar point groups, i.e. in groups that possess a polar axis \cite{Sodemann2015}. In contrast, it can be checked that the symmetric part transforms under any point group in exactly the same way as the gyration tensor \cite{yariv_optical_1984}, which is likewise a symmetric pseudotensor of rank two \cite{landau_electrodynamics_1984}.
Consequently, the symmetric part of $\db{D}$ can be nonzero only in point groups that admit a nonvanishing gyration tensor, i.e. in optically active groups \cite{landau_electrodynamics_1984,arteaga_optical_2014}. 
Therefore, crystallographic point groups that are neither polar nor optically active have a vanishing BD. \\

Among the 21 non-centrosymmetric crystallographic point groups, three ($\bar{4}3m$, $\bar{6}m2$ and $\bar{6}$) are neither polar nor optically active, further reducing the relevant candidates to 18. Moreover, within these, the two cubic chiral classes $432$ and $23$, support only a pure trace in the gyration tensor, while the BD tensor is inherently traceless. Therefore, only 16 of the 21 non-centrosymmetric point groups support a nonzero BD. These groups are: $1$, $2$, $3$, 
$4$, $6$, $222$, $32$, $422$, $622$, $m$, $mm2$, $3m$, $4mm$, $6mm$, $\overline{4}$ and $\overline{4}2m$. They include all polar and optically active point groups, except for the two cubic chiral groups mentioned above. Using the allowed BDs, it is possible to construct the electro-optic permittivity tensor for each crystallographic point group.

Following the classification diagram proposed by Halasyamani and Poeppelmeier \cite{Halasyamani1998}, non-centrosymmetric materials can be categorized according to four symmetry-dependent properties: polarity (spontaneous macroscopic electric polarization), optical activity (rotation of the plane of polarized light), piezoelectricity (generation of macroscopic electric polarization under mechanical stress) and enantiomorphism (existence of non-superimposable mirror-image crystal structures related to structural chirality). 
As we will demonstrate in the next subsections, this diagram also serves to classify the BD tensor of various materials according to their symmetry, which directly influences their non-conservative effects. As a result, it provides a valuable framework for organizing materials with different point-group symmetries based on their distinct non-conservative responses.

The diagram in Fig. \ref{fig1} organizes compounds by separating them into polar and nonpolar crystal sets, and then further classifying them according to properties such as optical activity, which correlates with circular dichroism, and piezoelectricity, which overlaps with materials that exhibit finite SHG.
As highlighted in Fig. \ref{fig1}, the 16 point groups capable of exhibiting a non-zero BD separate into five distinct categories (A-E).
First, among polar crystal classes, three non-overlapping categories can be defined: category A (crystal classes $1$, $2$, $3$, $4$, $6$), category B ($m$, $mm2$), and category C ($3m$, $4mm$, $6mm$).
Point groups within the same category (A-E) share a BD tensor with the same structure. 

Materials in category A exhibit all four symmetry-dependent properties: enantiomorphism, optical activity, pyroelectricity, and piezoelectricity, making them chiral, polar, optically active, and SHG-active. 
Category B materials are pyroelectric, piezoelectric, and optically active, while those in category C are pyroelectric and piezoelectric but lack optical activity.
Categories D, E include non-polar, non-centrosymmetric crystal classes. Category D materials are enantiomorphic, optically active, and piezoelectric, whereas category E materials are optically active and piezoelectric.

In this article, we will focus on specific symmetry categories shown in the diagram of Figure \ref{fig1} that are distinct from those examined in our previous studies \cite{mosfet, Rappoport2023, Morgado2024}.
Our objective here is not to provide a complete crystallographic classification, but rather to give a general picture of the symmetry-allowed electro-optic responses. We note in passing that our symmetry classification also applies to nonsymmorphic space groups, because the macroscopic form of the BD depends only on the linear point-group part of the space-group operations, and not on the fractional translations.\\
Furthermore, we will focus on point groups that possess a principal axis, with the static electric bias aligned parallel to this axis. Assuming that the principal axis is oriented along the $z$-direction, the BD tensor takes the generic form:
\begin{align} \label{eq:BDaxis}
\db{D}=\left(
\begin{array}{ccc}
 D_{xx} & D_{xy} & 0 \\
 D_{yx} & D_{yy} & 0 \\
 0 & 0 & D_{zz} \\
\end{array}
\right).
\end{align}
Table \ref{table2} shows the constraints on the BD tensor elements for the most relevant point groups. The explicit tensor forms for most point groups are provided in Appendix \ref{Dsymmetry}.

When the static bias is parallel to the principal axis ($\boldsymbol{\omega}_0 = \omega_{0z} \hat{\bf z}$), the non-Hermitian part of the electro-optic permittivity is governed by:
\begin{align} 
\label{eq:epsli2aux}
\db{\eps}_\text{EO}''=&
\frac{{\omega _{0z} }}{{\omega  + \tau ^2 \omega ^3 }}
\left(
\begin{array}{ccc}
 -D_{xy} & \frac{D_{xx}-D_{yy}}{2} & 0 \\
  \frac{D_{xx}-D_{yy}}{2} & D_{yx} & 0 \\
 0 & 0 & 0 \\
\end{array}
\right) + \nonumber \\
&
\frac{{\omega _{0z} \tau }}{{1  + \tau ^2 \omega ^2 }}
\left(
\begin{array}{ccc}
 0 & -i \frac{D_{xx}+D_{yy}}{2} & 0 \\
  i \frac{D_{xx}+D_{yy}}{2} & 0 & 0 \\
 0 & 0 & 0 \\
\end{array}
\right)
.
\end{align}
As seen, for the considered systems $\db{\eps}_\text{EO}''$ depends only on 4 elements of the BD tensor ($D_{ij}, \,\, i,j=x,y$).
The first tensor on the right-hand side of Eq. \eqref{eq:epsli2aux} corresponds to the component associated with linear dichroic gain ($\db{\eps}_\text{EO,R}''$), while the second term represents the component associated with chiral gain ($\db{\eps}_\text{EO,I}''$).

Consistent with the analysis in Sec. \ref{Sec:EOgain}, the tensor $\db{\eps}_\text{EO}''$ has a trivial eigenvalue ($\lambda_3 = 0$). Consequently, the NHEO response is determined by two orthogonal eigenpolarizations, ${\bf{u}}_1$ and ${\bf{u}}_2$, which lie in the $xoy$ plane, perpendicular to the bias field ${\bf{E}}_0$. Notably, the chiral-gain component of the response appears only for point groups where $D_{xx} + D_{yy}$ can be nontrivial. Furthermore, the trace of $\db{\eps}_\text{EO}''$, which equals the sum of the eigenvalues, is proportional to $D_{yx}-D_{xy}$. Therefore, when $D_{yx}=D_{xy}$, the response becomes indefinite, and $\lambda_2=-\lambda_1$.

\begin{table*}[th!]
\centering
\begin{tabular}{ |c|c|c| } 
 \hline
 {\bf Point group} & {\bf Constraints on $\db{D}$} & {\bf Non-Hermitian response} \\
 \hline
 {$3, \,\, 4, \,\, 6$ (Cat. A) } & $D_{xx}=D_{yy}=-D_{zz}/2 \equiv D_0, \,\, D_{yx}=-D_{xy}$ & Both components  \\ 
  \hline
 {$mm2$ (Cat. B)} &  $D_{xx}=D_{yy}=D_{zz}=0$ & Linear dichroic gain \\ 
  \hline
 {$3m, \,\, 4mm, \,\, 6mm$ (Cat. C)} & $D_{xx}=D_{yy}=D_{zz}=0, \,\, D_{yx}=-D_{xy}$  & Linear dichroic gain \\ 
  \hline
 {$32, \,\, 422, \,\, 622$ (Cat. D)} & $D_{xx}=D_{yy}=-D_{zz}/2 \equiv D_0, \,\, D_{yx}=D_{xy}=0$ & Chiral gain \\ 
  \hline
  {$222$ (Cat. D)} &   $D_{yx}=D_{xy}=0$ & Both components \\ 
  \hline
 {$\overline{4}$ (Cat. E)} & $D_{xx}=-D_{yy} \equiv D_0, \,\, D_{yx}=D_{xy}, \,\, D_{zz}=0$   & Linear dichroic gain\\ 
 \hline
 {$\overline{4}2m$ (Cat. E)} & $D_{xx}=-D_{yy} \equiv D_0, \,\, D_{yx}=D_{xy}= D_{zz}=0$   & Linear dichroic gain\\ 
 \hline
\end{tabular}
\caption{Constraints on the BD tensor and the type of NHEO response corresponding to each point group. The material is assumed to have a symmetry axis along the $z$-direction.}
\label{table2}
\end{table*}

In the next subsections, we discuss the non-conservative response of each category.
For the sake of simplicity, we discuss only the point groups that have up to two independent parameters. According to our previous discussion, it excludes point groups $1$, $2$ and $m$.

\subsection{Pure linear dichroic gain: categories B, C and E} \label{sec:cat_BCE}
One of the most interesting scenarios originates from the BD of the point group $mm2$, which belongs to category B.
For this group, considering a polar axis directed along $z$, the BD has the form in Eq. \eqref{eq:BDaxis} with all diagonal elements equal to zero. 

As can be seen in Eq. \eqref{eq:epsli2aux}, for an electric bias applied along the polar axis,  $\db{\eps}_\text{EO}''$ is real-valued and diagonal for this point group ($D_{xx}=D_{yy}=0$). Thus, the material exhibits a reciprocal response characterized by linear dichroic gain. The nontrivial eigenvalues $\lambda_1$ and $\lambda_2$ of $\db{\eps}_\text{EO}''$ are given by
\begin{align} 
\lambda_{1}=-\frac{  D_{xy} \omega_{0z}}{\tau ^2 \omega ^3 +\omega  }, \quad \lambda_{2}=\frac{  D_{yx} \omega_{0z}}{\tau ^2 \omega ^3 +\omega  }.
\end{align}
The corresponding eigenvectors are linearly polarized along ${\bf{u}}_1=\hat{\vec{x}}$ and along ${\bf{u}}_2=\hat{\vec{y}}$, respectively. 
The signs of the eigenvalues are determined by the signs of $D_{yx}$ and $D_{xy}$, which can either be the same or opposite. This leads to two distinct scenarios: in the first, the system exhibits gain for one polarization and loss for the orthogonal polarization (referred to as indefinite gain). In the second scenario, the system exhibits gain (or dissipation) for all polarizations in the $xoy$ plane, which can be switched to loss (or gain) by simply reversing the direction of the electric bias.

\begin{figure}
	
		\includegraphics[width=0.95\linewidth]{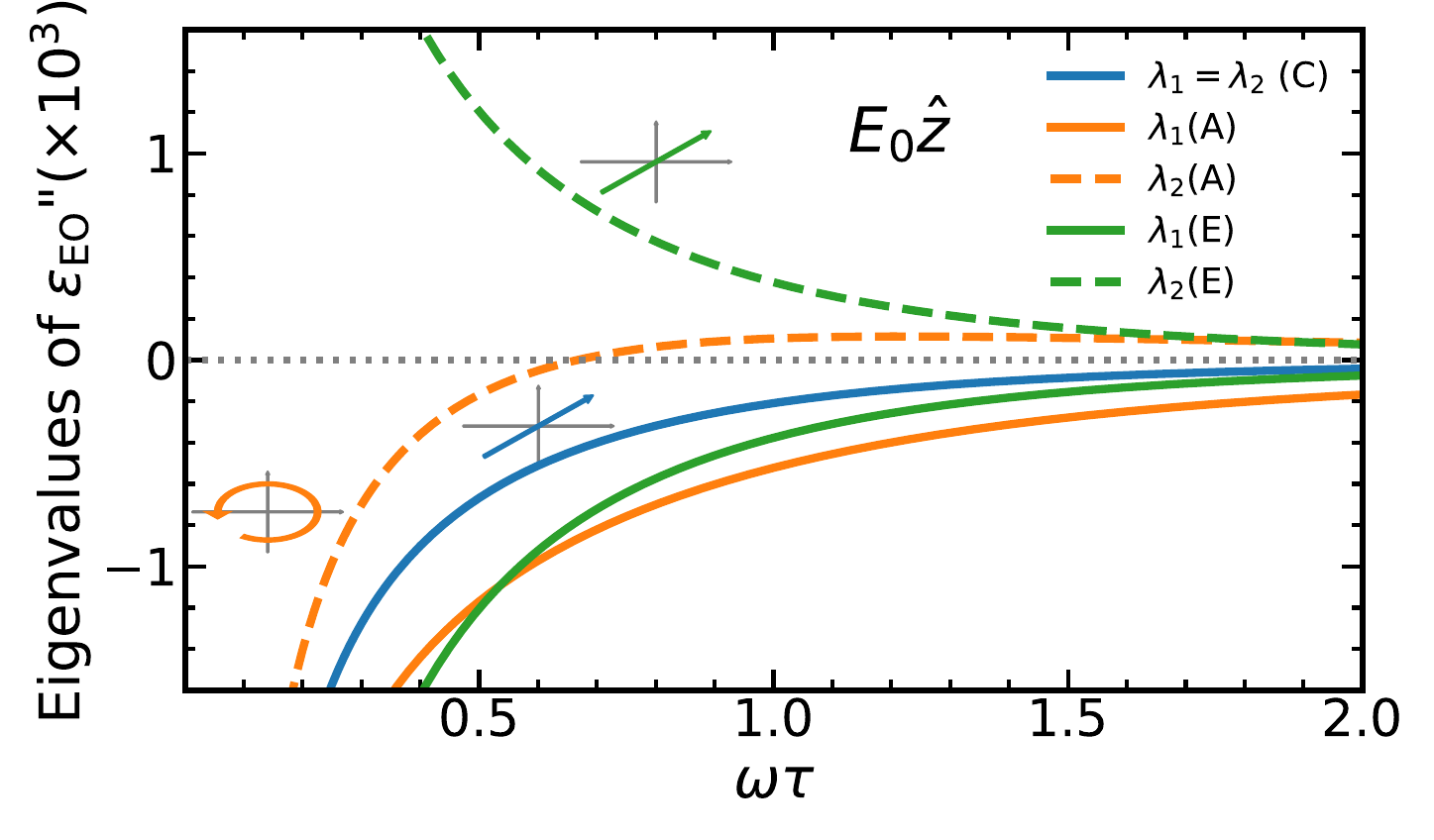}\caption{Eigenvalues $\lambda_1$ and $\lambda_2$ of the non-conservative part of the electro-optic permittivity tensor $\db{\eps}_\text{EO}''$ for an electric bias along the principal axis and different point group symmetries belonging to the different categories (the category is indicated by a letter in parentheses). Negative eigenvalues indicate gain whereas positive eigenvalues correspond to loss. The insets illustrate the eigenpolarizations  of $\db{\eps}_\text{EO}''$.
        The numerical parameters are $D_{xy}=1$, $D_0=1.5$, $\tau=1\,\rm{ps}$ and $\omega_{0z}/{(2\pi)}\approx 66.47 \,\rm{THz}$ ($E_0=10^4$ V/m). }\label{fig2}
		
\end{figure}

Point groups belonging to category C  represent a special case of the BD tensor for the group $mm2$, where the BD is purely antisymmetric: $D_{yx} = -D_{xy}$. These materials are polar but do not present any optical activity. The nontrivial eigenvalues $\lambda_1$ and $\lambda_2$ of $\db{\eps}_\text{EO}''$ are doubly degenerate and given by 
\begin{equation}
\lambda_1 = \lambda_2 = -  D_{xy} \omega_{0z}/(\tau ^2 \omega ^3 +\omega).
\end{equation}
Therefore, for the point groups belonging to category C, all polarizations in the $xoy$ plane result in either gain or absorption, depending on the sign of $E_{0z}$. This behavior is illustrated by the solid blue curve in Fig. \ref{fig2}, which considers an electric bias that results in gain. 

This response underscores the critical importance of the symmetry properties of the material and their impact on the NHEO response. When $\db{\eps}_\text{EO}''$ is real and symmetric, the system has a reciprocal (non-gyrotropic) response with optical gain. This feature is particularly relevant for applications requiring polarization-independent optical gain, while still allowing the gain to be controlled via the applied bias.

We emphasize that when an electric bias is applied along the polar axis, the discussed groups do not exhibit any electro-optic-induced nonreciprocal gyrotropic response. This is because, as previously noted, a gyrotropic response always coexists with a chiral-gain response, which is absent in the considered point groups.

Materials from category E also lead to pure linear dichroic gain. For the group $\overline{4}$, the BD tensor is such that $D_{xx}=-D_{yy}\equiv D_0$ and $D_{xy}=D_{yx}$ (see Table \ref{table2}).
Now, the eigenvalues of $\db{\eps}_\text{EO}''$ for eigenpolarizations in the $xoy$ plane are 
\begin{align} 
\lambda_\pm&=\pm\frac{\omega_{0z} \sqrt{D_0^2+D_{xy}^2}}{\tau ^2 \omega ^3+\omega }.
\end{align}
The associated eigenvectors are independent of frequency and are linearly polarized:
\begin{align} 
\vec{E}_\pm \sim \left(-D_{xy} \pm \sqrt{D_0^2 + D_{xy}^2}\right) \hat{\vec{x}} + D_0 \hat{\vec{y}}.
\end{align}
Hence, these materials may behave as indefinite gain media with the gain and dissipation activated by orthogonal linear polarizations. 
The response in the \(\overline{4}2m\) point group represents a particular case of that in the \(\overline{4}\) group where \(D_{xy} = 0\). 
In this case, the two eigenvalues display a frequency dependence that is  symmetric about the frequency axis, as illustrated by the green curves in Fig. \ref{fig2}.

\subsection{Category D} 
Next, we analyze the point groups in category D, where the BD is diagonal and satisfies $D_{zz} = -D_{xx} - D_{yy}$.

For the point groups $32$, $422$, and $622$, the BD tensor is further constrained by the condition $D_{xx} = D_{yy} \equiv D_0$. From Eq. \eqref{eq:epsli2aux}, it follows that these point groups give rise to a pure chiral-gain response, with the chiral gain vector aligned along the material's symmetry axis ($\boldsymbol{\Omega}_\omega \sim \hat{\bf{z}}$). As a result, the eigenvectors are frequency-independent and exhibit circular polarization:
\begin{align} \vec{E}_\pm \sim \frac{1}{\sqrt{2}} \left(\pm i \hat{\vec{x}} + \hat{\vec{y}} \right). \end{align}
The corresponding eigenvalues are
$ \lambda_\pm= \pm \frac{ \omega_{0z} \tau  D_0}{ 1 + \tau ^2 \omega ^2}$.
Additionally, $\db{\eps}_\text{EO}''$ is purely imaginary, which significantly enhances the gyrotropic effects. Specifically, for an arbitrarily linearly polarized wave propagating along the $z$-direction, EO-induced gyrotropy enhances the polarization rotation within the material (the kinetic Faraday effect).

For a chiral-gain response, one circular polarization state undergoes amplification, while the other experiences attenuation \cite{lannebere_chiral_2025}. The handedness of the polarization experiencing gain is determined by the orientation of the bias along the principal axis. This property makes chiral gain a powerful tool for achieving controlled polarization dynamics and enhanced chiral selectivity within the material.

The ability to control and generate circularly polarized light using chiral gain opens the door to innovative electromagnetic devices with unique functionalities. These include chiral lasers, polarization-dependent amplifying mirrors, and loss-compensated photonic waveguides \cite{Hakimi2024,lannebere_chiral_2025,Serra2024,Prudencio2024}. Moreover, chiral gain may contribute to enhancing the sensitivity of chiral sensing and spectroscopy.

For the $222$ group, which also belongs to category D but has $D_{xx} \ne D_{yy}$, the nontrivial eigenvalues of $\db{\eps}_\text{EO}''$ are given by
\begin{align} 
\lambda_\pm= \pm \frac{ \omega_{0z} \sqrt{\tau ^2 \omega ^2 (D_{xx}+D_{yy})^2+(D_{xx}-D_{yy})^2}}{2 \left(\tau ^2 \omega ^3+\omega \right)},
\end{align}
indicating that the gain remains indefinite. The corresponding eigenpolarizations are  
\begin{align} 
\vec{E}_\pm \sim \pm  \frac{ (D_{xx}-D_{yy})-i\tau  \omega  (D_{xx}+D_{yy})}{\sqrt{\tau ^2 \omega ^2 (D_{xx}+D_{yy})^2+(D_{xx}-D_{yy})^2}}   \hat{\vec{x}} +\hat{\vec{y}}.
\end{align}
Unlike the previously discussed cases, the eigenpolarizations in this scenario are frequency-dependent and exhibit elliptical polarization. This behavior arises because, for this point group, the NHEO response described by Eq. \eqref{eq:epsli2aux} (with $D_{xx} \ne D_{yy}$ and $D_{xy} = D_{yx} = 0$) incorporates both chiral-gain and linear dichroic components.

\subsection{Category A} \label{sec:cat_A}

For point groups in category A (specifically groups 3, 4 and 6), the BD is 
subject to the constraints $D_{xx}=D_{yy}=-D_{zz}/2 \equiv D_0$ and $D_{yx}=-D_{xy}$ (see Table \ref{table2}).
For a bias along the polar axis, the nontrivial eigenvalues of $\db{\eps}_\text{EO}''$ are 
\begin{align} 
\lambda_1=-\frac{ \omega_{0z} \left(D_{xy}+ D_0  \tau  \omega  \right)}{\tau ^2 \omega ^3 +\omega }, \quad \lambda_2=-\frac{ \omega_{0z} \left(D_{xy}  - D_0   \tau  \omega \right) }{\tau ^2 \omega ^3 +\omega }, 
\end{align}
where negative eigenvalues indicate gain, whereas positive eigenvalues correspond to loss. This implies that, depending on the sign of $\frac{D_0\tau  \omega +  D_{xy}}{D_0\tau  \omega -  D_{xy}}$, the medium can exhibit either a definite or indefinite non-conservative response across all polarizations with electric field in the $xoy$ plane. 

As the tensor $\db{\eps}_\text{EO}''$ contains both linear dichroic and chiral-gain components for this point group, the specific behavior is frequency-dependent. This property enables a transition from an indefinite non-conservative response to a semi-definite (positive or negative) response, where all field polarizations are either amplified or absorbed, depending on the frequency. This transition occurs at the critical frequency $\omega_c=\left|D_{xy}/(\tau D_0)\right|$ for which the linear dichroic gain and the chiral-gain have equal strength.  
In the example shown by the orange curves in Fig. \ref{fig2}, $\omega_c \tau \approx 0.67$.

At low frequencies ($\omega\tau \rightarrow 0$), the two eigenvalues become degenerate, given by $\lambda_1 = \lambda_2 = -\omega_{0z} D_{xy}/\omega$. In this regime, the linear dichroic gain or dissipation component dominates, leading to either uniform gain or absorption across all polarizations. In contrast, for $\omega>\omega_c$, the chiral-gain component of the non-conservative response becomes dominant, rendering the response indefinite.

Interestingly, even though $\db{\eps}_\text{EO}''$ [see Eq. \eqref{eq:epsli2aux}] has in general both real and imaginary components, its eigenvectors are always circularly polarized when $D_0 \neq 0$:
$\vec{E}_1 \sim \left(i\hat{\vec{x}} + \hat{\vec{y}}\right)/\sqrt{2}$ and $\vec{E}_2 \sim \left(-i\hat{\vec{x}} + \hat{\vec{y}}\right)/\sqrt{2}$. The justification for this property is that the linear dichroic gain component $\db{\eps}_\text{EO,R}''$ is a scalar in the $xoy$ plane. 
Furthermore, the presence of a chiral-gain response inherently implies that  \(\db{\eps}_\text{EO}'\) 
induces a nonreciprocal gyrotropic response. 

\subsection{Biases along other crystallographic directions} 

Up to this point, we have focused exclusively on an electric bias along the $z$-axis, which for polar classes coincided with the polar axis. However, by applying the electric bias along another direction, one may further expand the range of achievable optical responses. Here, we summarize the key findings.

\begin{figure}
	
		\includegraphics[width=0.95\linewidth]{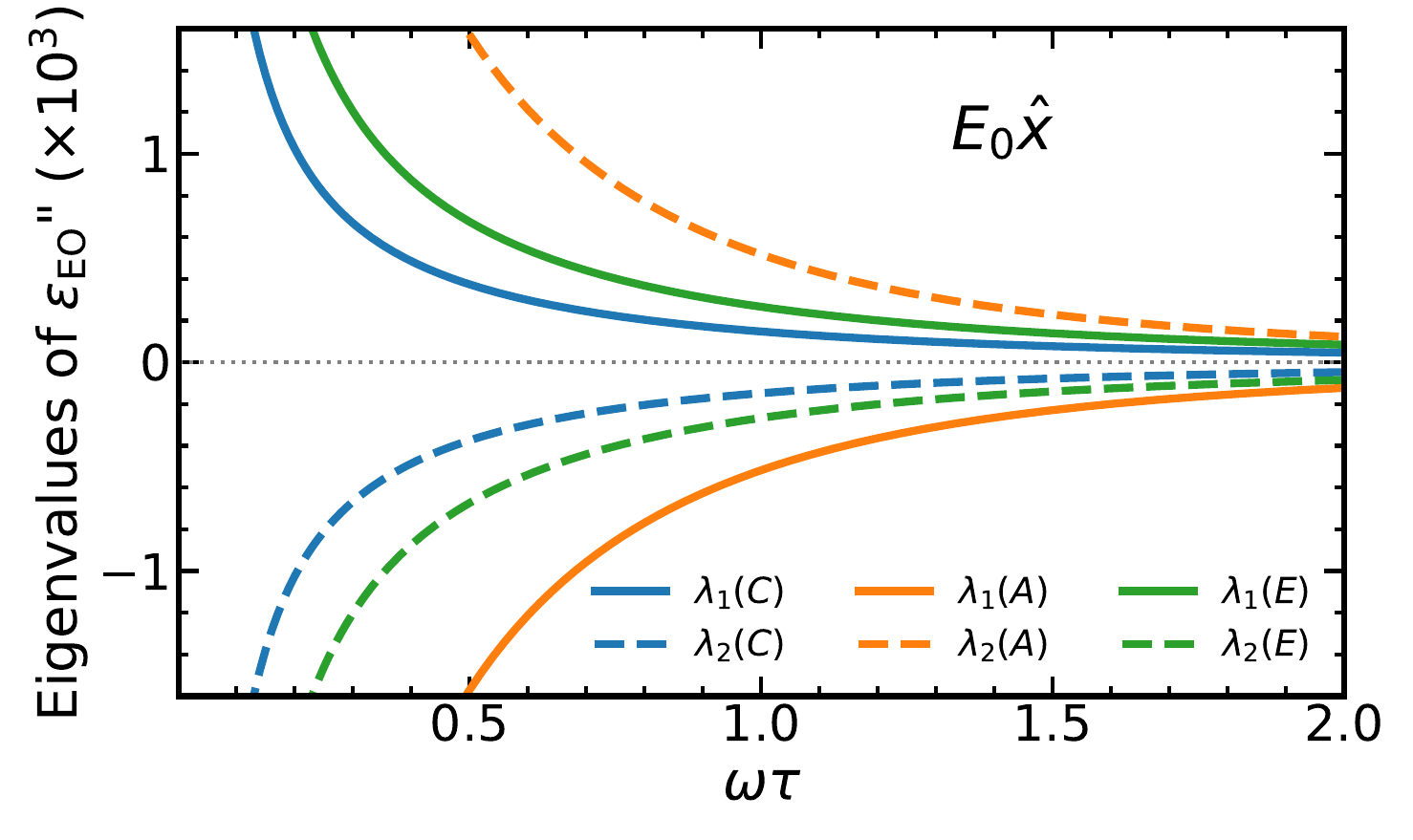}\caption{Eigenvalues $\lambda_1$ and $\lambda_2$ of the non-Hermitian part of the electro-optic permittivity tensor $\db{\eps}_\text{EO}''$ for an electric bias along the $x$-axis and different categories of crystallographic point group symmetries. The numerical parameters are the same as those used in Fig. \ref{fig2}.}\label{fig3}
		
\end{figure}

For an electric bias along a generic crystallographic direction, the tensor $\db{\eps}_\text{EO}''$ typically  has both linear gain and chiral-gain components. Thereby, its eigenvectors and eigenvalues are typically frequency dependent with the eigenpolarizations  elliptically polarized. Moreover, since the nonreciprocal gyrotropic response is intrinsically linked to the chiral-gain response, it also manifests in most cases.

For a generic bias direction,
most point groups exhibit three non-trivial eigenvalues $\lambda_i$. As $\db{\eps}_\text{EO}''$  cannot be a definite tensor, either one eigenvalue is positive and the other two are negative, or two eigenvalues are positive and one is negative. An exception occurs for the groups $32$, $422$, and $622$ (category D), where one eigenvalue remains zero, and the other two satisfy $\lambda_1 = -\lambda_2$. 

For a horizontal bias (relative to the symmetry axis), it can be shown that only the elements $\eps_{xz}''$, $\eps_{yz}''$, $\eps_{zx}''$, and $\eps_{zy}''$  are nonzero. In particular, the trace of \(\db{\eps}_\text{EO}''\) is zero, implying that the sum of all eigenvalues must also be zero. Moreover, the determinant of \(\db{\eps}_\text{EO}''\) is identically zero for this type of matrix structure, meaning that one of its eigenvalues necessarily vanishes, while the remaining two are equal in magnitude but opposite in sign. In general,   $\db{\eps}_\text{EO}''$ has both chiral-gain and linear-dichroic gain components, resulting in elliptical eigenpolarizations that depend on frequency. 

Figure \ref{fig3} presents the two non-zero eigenvalues, $\lambda_1$ and $\lambda_2$, of the nonconservative part of the electro-optic permittivity tensor, $\db{\eps}_\text{EO}''$, as a function of frequency, for various crystal symmetry categories under an electric bias along the $x$-axis. In agreement with the previous discussion,  the eigenvalues exhibit symmetry with respect to the frequency axis across all considered symmetry categories, indicating that one eigenvalue corresponds to gain while the other corresponds to loss. Consequently, for a horizontal bias, the electro-optic response is always characterized by indefinite gain. As before, the gain and loss responses are interchanged when the electric bias is reversed.

It is relevant to note that when the electric bias is aligned with the symmetry axis of the material (as in Sects. \ref{sec:cat_BCE} to \ref{sec:cat_A}), the electro-optic response is insensitive to $D_{zz}$. When $D_{zz}$  is the dominant element of the BD tensor, an oblique bias can often provide larger optical gain. Furthermore, numerical simulations show that even when $D_{zz}$  is not dominant, a horizontal bias may enhance the gain for some material systems.

\section{Criteria for Optical Gain} \label{sec:criteria}

So far, we have focused solely on the linear electro-optic response of 3D materials with nontrivial BDs. However, to fully understand the conditions for optical gain, it is important to also consider the material's response in the absence of an electric bias. In particular, since these materials must exhibit a metallic response with electronic scattering \cite{Morgado2024}, we should take into account the linear Drude-type response, which describes the behavior of free electrons. In this context, we assume for simplicity that the material response in the absence of bias is characterized by an isotropic Drude model. This allows us to establish a straightforward criterion for the emergence of optical gain. The  permittivity of Drude's model is given by $\eps_\text{m}=\eps_\text{m}'+i\eps_\text{m}''$, where the real $\eps_\text{m}'$ and imaginary $\eps_\text{m}''$ parts are given by
\begin{equation}
 \eps_\text{m}'=1-\frac{ \omega_p^2 \tau^2   }{  \omega^2 \tau^2 +1  },~~~
 \eps_\text{m}''=  \frac{ \omega_p^2   \tau  }{ \omega^3 \tau^2 +\omega },
\end{equation}
and $\omega_p$ is the plasma frequency. The high-frequency dielectric constant was chosen to be equal to one. At low frequencies $\omega \to 0$, the Drude losses tend to  $\eps_\text{m}'' \to   \omega_p^2   \tau  / \omega$  and have the same frequency dependence $1/\omega$  as the EO susceptibility. \\ 
The full permittivity response includes both the Drude term and the linear electro-optic term:
\begin{align}
\db{\eps} = \eps_\text{m} \vec{1}_{3\times3}+ \db{\eps}_\text{EO} 
 = \db{\eps}'+ i\db{\eps}''
 \end{align}
 with $\db{\eps}'=\eps_\text{m}' \vec{1}_{3\times3} + \db{\eps}_\text{EO}'$ denoting
 the Hermitian part of the response while  $\db{\eps}''=\eps_\text{m}'' \vec{1}_{3\times3} + \db{\eps}_\text{EO}''$ governs the non-Hermitian part, which determines the exchange of power between the material and the wave [see Eq. \eqref{E:powerdip}].
 
 As described in previous sections, the sign of the eigenvalues of $\db{\eps}''$ determines whether the system exhibits gain or loss. The eigenvalues of $\db{\eps}''$ are given by $\lambda_i + \eps_\text{m}''$, where $\lambda_i$ ($i=1,2,3$) are the eigenvalues of $\db{\eps}_\text{EO}''$. Thus, the threshold for optical gain can be found by solving $\lambda_{i} + \eps_\text{m}''=0$. The eigenvectors of $\db{\eps}''$, in turn, are identical to those of  $\db{\eps}_\text{EO}''$.

Focusing on a bias along the $z$-axis, we can explicitly write the gain criteria for two sets of interesting symmetries. For category C, gain occurs when $\omega_{0z}D_{xy} > \omega_p^2 \tau$. For category A, the two different eigenvalues provide the conditions $\omega_{0z} \left(D_0 \tau \omega + D_{xy}\right) > \omega_p^2 \tau$ and $\omega_{0z} \left(D_{xy} - D_0 \tau \omega\right) > \omega_p^2 \tau$. Both conditions can be satisfied simultaneously, and in that case, the two modes provide gain. At low frequencies, where $\omega \tau \to 0$, both conditions reduce to the same equation. In this limit, we can achieve gain for both polarizations when $\omega_{0z}D_{xy} > \omega_p^2{\tau}$. Although the threshold expressed in terms of $\omega_{0z}$ is proportional to $\tau$, $\omega_{0z}$ itself is proportional to $\tau E_0$. Consequently, the threshold electric bias is independent of $\tau$ within the present single-relaxation-time Drude model.

For nanophotonic applications, the most relevant regime is typically the one in which the operating frequency $\omega$ is comparable to or larger than the plasma frequency $\omega_p$, or more generally the epsilon-near-zero (ENZ) frequency, since the material becomes transparent in this range. Due to this reason, gain effects are most relevant when the bias strength $\omega_0$ is comparable to $\omega_p$. Consequently, systems in which the effective plasma frequency (ENZ point) is in the mid infrared or optical ranges require, in practice, a stronger bias to achieve gain than systems where the ENZ point lies in the terahertz range.


\section{Candidate Materials}

The symmetries required for the non-Hermitian linear electro-optic effect are typically found in materials such as borates \cite{mutailipu_borates_2021}, chiral perovskites \cite{long_chiral-perovskite_2020}, and transition metal dichalcogenides. However, for this effect to be significant, the materials must exhibit large BDs and low effective plasma frequency \cite{Morgado2024}. In this regard, Weyl semimetals (WSMs) stand out as strong candidates due to their intrinsic topological properties \cite{ruan_ideal_2016,qian_weyl_2020}, which naturally give rise to large BD \cite{facio_strongly_2018,zhang_berry_2018}. Many of these WSMs fall under categories B and C, making them ideal for observing the linear dichroic gain.

For instance, WSMs such as TaAs, TaP, NbAs, NbP and In-doped Pb$_{1-x}$Sn$_x$Te belong to the $4mm$ point group. These materials exhibit large BD \cite{zhang_berry_2018,zhang_giant_2022} and show electro-optic responses classified under category C. On the other hand, WSMs such as MoTe$_2$, WTe$_2$, and NbIrTe$_4$  fall under the point group $mm2$ \cite{zhang_berry_2018,sharma_room-temperature_2019,ma_growth_2022,nishijima_ferroic_2023,lee_spin-orbit-splitting-driven_2024} and exhibit electro-optic responses associated with category B.

In addition, materials with other symmetries are also promising candidates. For example, BiTeI, which belongs to the $3m$ symmetry group of category C, and WSM in chalcopyrites $\mathrm{CuTlSe}_2$, $\mathrm{AgTlTe}_2$, $\mathrm{AuTlTe}_2$, and $\mathrm{ZnPbAs}_2$, which belong to the $\overline{4}2m$ symmetry group and are classified in category E \cite{ruan_ideal_2016}. Other examples are WSMs with $\bar{4}$ symmetry \cite{Gao2021}, which belong to category E. In category D, tellurium has been explored in the context of NHEO responses \cite{Morgado2024}.
Currently, there is a scarcity of reported materials in category A, and their non-linear characteristics remain underexplored. The numerical values of the BD used in Figs. \ref{fig2} and \ref{fig3} are consistent with the order of magnitude reported in the literature, which in some cases reaches values as large as 20 \cite{zhang_berry_2018}.

\section{Reflectance Analysis}

Next, we study the impact of the different NHEO permittivity tensors on electromagnetic wave scattering. Specifically, we investigate the scattering of a plane wave by a low-symmetry metallic material slab that is backed by a reflective mirror.  Such non-conservative mirrors can either amplify or attenuate waves based on light polarization and electric bias direction. 

We consider a slab of material of thickness $d$ terminated by a perfect electric conducting (PEC) wall. A plane wave propagates along the $+z$-direction and impinges on the slab, at $z=0$, along the normal direction. Let $ \vec{E}_t^\text{inc}=E_x^\text{inc} \hat{\vec{x}}  + E_y^\text{inc} \hat{\vec{y}}$ be the complex amplitude of the (transverse) incident field and $ \vec{E}_t^\text{ref}$ the complex amplitude of the (transverse) reflected field. At the $z=0$ interface  these fields are related by
 \begin{align}\label{E:E_reflected}
\vec{E}_t^\text{ref}(z=0)&=\db{\vec{R}}\cdot \vec{E}_t^\text{inc}(z=0)
\end{align}
where $\db{\vec{R}}$ is the 2$\times$2 reflection matrix. $\db{\vec{R}}$ can be calculated using transfer matrix methods. A related calculation can be found in Ref. \onlinecite{lannebere_chiral_2025} and the detailed analysis for this setup is discussed in Appendices \ref{AppendR} and \ref{AppendRr}. The eigenvectors of the (Hermitian nonnegative) reflectance matrix $\db{\cal R}=\db{\vec{R}}^\dagger\cdot \db{\vec{R}}$ determine the polarizations of the incident wave that maximize and minimize the reflected power \cite{lannebere_chiral_2025}. 

\begin{figure}
	
		\includegraphics[width=0.98\linewidth]{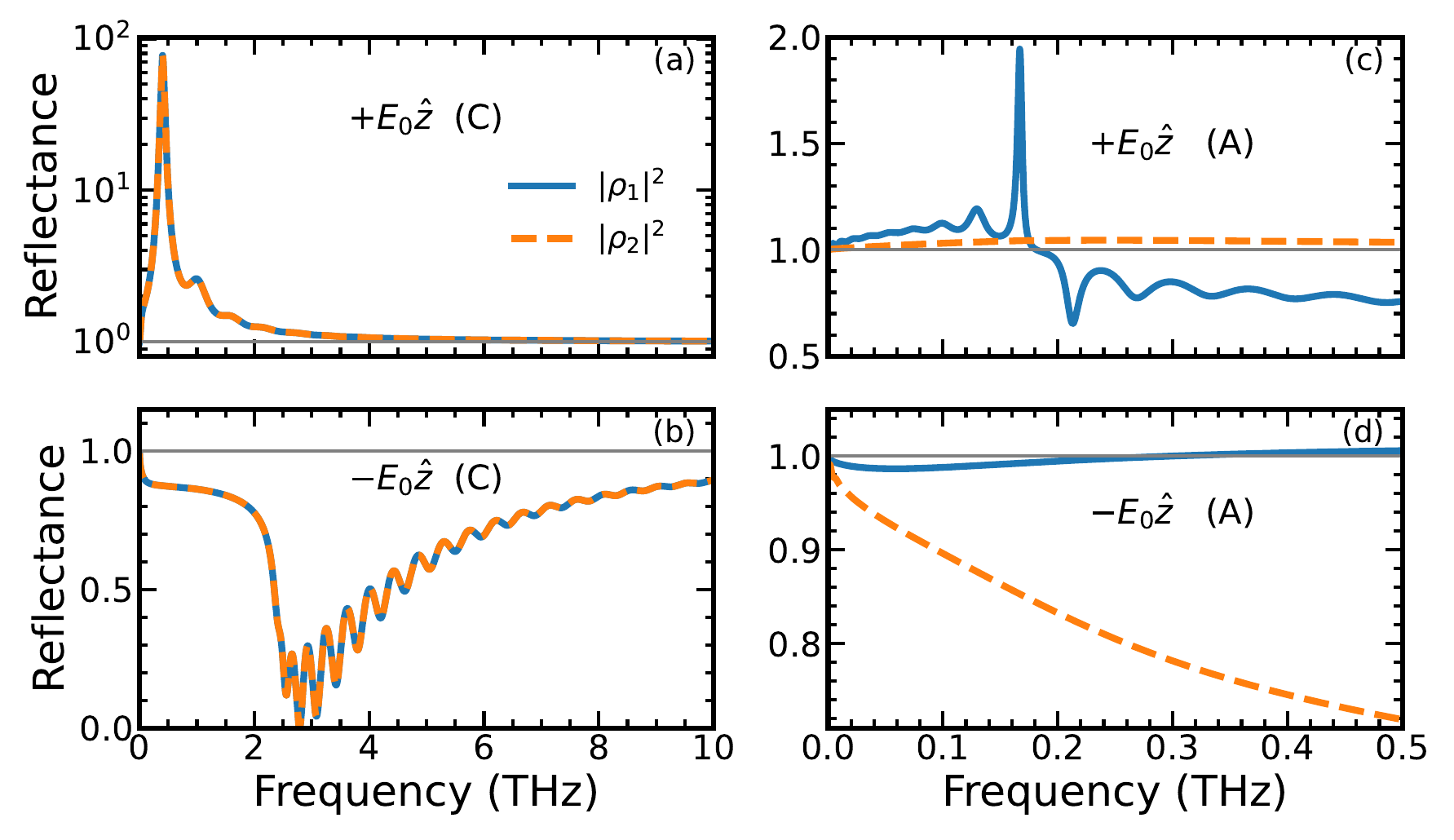}\caption{Eigenvalues of the reflectance matrix as a function of frequency for different point groups. The left panels show results for category C while the right panels depict results for category A. On the top (bottom) panels, the materials have a positive (negative) electric bias along the $z$-axis. The parameters used in these calculations are: dipole components $D_{xy} = 1$ and $D_{0} = 2/3$, transport relaxation time $\tau$ = 1 ps, plasma frequency $\omega_p/{(2\pi)}\approx 1.59 \,\rm{THz}$, and material thickness of $d = 300 \mu$m. For the left panels, $E_0 = 2873.1$ V/m ($\omega_{0z}/{(2\pi)}\approx 19.1 \,\rm{THz}$), and for the right panels, $E_0 = 10^4$ V/m ($\omega_{0z}/{(2\pi)}\approx 66.47 \,\rm{THz}$).\label{fig4}}.

\end{figure}

Figure \ref{fig4} shows the reflectance as a function of frequency for various point-group symmetries. The left panels display results for category C while the right panels show results for category A. The top of the figure represents materials with positive electric bias along the $z$-axis, whereas the bottom panels corresponds to materials with a negative bias. This arrangement allows for a clear comparison of how the change in bias direction affects the reflectance across different frequencies and point-group symmetries. We account for the effects of losses originating from the Drude model of optical conductivity. In all panels, we used a transport relaxation time of $\tau$ = 1 ps and a plasma frequency of $\omega_p/{(2\pi)}\approx 1.59 \,\rm{THz}$. As discussed in section \ref{sec:criteria}, a relatively low plasma frequency may be beneficial for achieving optical gain. 

The material slab has a thickness of $d = 300 \mu$m. In the left panels, we assume an electric field bias of $E_0 = 2873.1$ V/m, whereas in the right panels we use $E_0 = 10^4$ V/m. In both cases, the bias strength was chosen according to the gain criteria discussed at the end of section \ref{sec:criteria}. For the two sets of point groups, we consider $D_{xy} = 1$, and for the point groups $3$, $4$, and $6$ (category A), we also use $D_0 = \frac{2}{3}$. 
The values of $D_{xy}$ and $D_{0}$ are consistent with the literature. Still, some Weyl semimetals such as NbP (category C), have estimated $D_{xy}$ of the order of 20 \cite{zhang_berry_2018}, which could make it possible to have equivalent optical gains with electric biases of the order of $E_0 = 10^2$ V/m.\\
It is however important to keep in mind that materials exhibiting large dipoles are often characterized by narrow electronic band gaps, which correspond to a lower threshold for interband effects. These interband effects introduce additional contributions to both the linear and electro-optic components of the conductivity, which are not captured by our theoretical model.

As seen in Figure \ref{fig4}, for category C, the two eigenvalues of $\db{\cal R}$ result in identical reflectance,  which means that the amplification or attenuation of the incident wave is independent of the polarization. With a positive bias, there is a significant gain at low frequencies (reflectance is greater than unity), while a negative bias leads to strong absorption (small reflectance) at intermediate frequencies.

In contrast, for category A, the two eigenvalues produce different reflectances. With a positive bias, both eigenvalues show gain at low frequencies, but as the frequency increases, one eigenvalue continues to produce gain while the other experiences losses. For a negative bias, both eigenvalues produce losses at low frequencies. However, as the frequency increases, a transition occurs, leading to one of the eigenvalues generating gain. This behavior arises because materials in category A have both chiral-gain and linear-dichroic gain components, with the linear-dichroic gain dominating at low frequencies and the chiral-gain component prevailing at high frequencies.

The gain can be further enhanced by optimizing the slab thickness (not shown). For category C, increasing the thickness up to $d = 1000 \mu$m can boost reflectance by two orders of magnitude or more. In contrast, for category A, reducing the thickness to $d = 50-100 \mu$m can enhance reflectance by up to a factor of ten. In both cases there is a single optimal value for the thickness that maximizes the reflectance.


We note that the static electric fields considered here, 10$^3$-10$^4$ V/m, are within experimentally accessible ranges used in nonlinear Hall and BD measurements \cite{Ye2023}. These field strengths are well below typical dielectric breakdown thresholds of thin-film and two-dimensional devices \cite{ranjan_dielectric_2021}, and Joule heating can be further minimized under low-duty-cycle or pulsed excitation.

\section{Conclusions}

We investigated the non-Hermitian linear electro-optic effect in crystals, focusing on how different point group symmetries influence the resulting optical response. 
Through a detailed symmetry analysis of the 32 crystallographic point groups, we have outlined a comprehensive roadmap for engineering non-reciprocal optical responses with gain for specific light polarizations, with potential applications in optical devices.

Our analysis shows that the NHEO response of a material is intricately linked to its point-group symmetry. We have organized the non-centrosymmetric crystal materials into several symmetry-based categories (A, B, C, D, and E) and explored the distinctive optical properties each category exhibits. Materials in categories B, C and E demonstrate linear dichroic gain responses when the electric bias is aligned with the principal axis. Notably, the eigenpolarizations that activate the gain or loss are linearly polarized, and the gain or dissipation response can be reversed by simply changing the direction of the applied static bias. These materials provide a reciprocal gain response free from gyrotropic effects. For certain point groups, it is possible to achieve gain (or dissipation) for all polarizations in the plane perpendicular to the applied bias, while for others, the response remains indefinite.

In contrast, materials belonging to category D exhibit chiral optical gain, where light of one handedness is amplified while light of opposite handedness is absorbed, showcasing the potential for chiral-selective applications. Furthermore, in these materials the electro-optic effect also induces a nonreciprocal gyrotropic response. Additionally, materials in category A typically feature both linear dichroic and chiral-gain components. These materials may undergo a transition that determines which component dominates.

Weyl semimetals emerge as particularly strong candidates for exhibiting prominent NHEO effects due to their large BDs. Specifically, many Weyl semimetals belong to the $mm2$ and $4mm$ point group symmetries (e.g., NbP, TaAs, NbAs, and MoTe$_2$), making them ideal for observing NHEO effects and for tuning linear dichroic optical gain in response to external biases. 

We also presented a reflectance analysis that illustrates the practical implications of the NHEO effect. In particular, we demonstrated how wave amplification or attenuation can be achieved using non-Hermitian mirrors, utilizing materials with different symmetries through either linear dichroic gain or chiral-gain. Our analysis provides important insights into how the NHEO effect could be harnessed in practical applications, such as in the design of non-reciprocal optical devices or systems where polarization-dependent optical gain is desirable.
\begin{acknowledgments}
This work was partially funded by the Institution of Engineering and Technology (IET), by the Simons Foundation (Award SFI-MPS-EWP-00008530-10) and by FCT/MECI through national funds and when applicable co-funded EU funds under UID/50008: Instituto de Telecomunicações.
SL acknowledges FCT and IT-Coimbra for the research financial support with reference DL 57/2016/CP1353/CT000.
TGR acknowledges financial support from FAPERJ,  grant numbers E-26/200.959/2022 and E-26/210.100/2023, CNPq, Grant  403130/2021-2, FCT - Fundação para a Ciência e Tecnologia,  project reference numbers UIDB/04650/2020, 2022.06797.PTDC and 2022.07471.CEECIND/CP1718/CT0001
(with DOI identifier: 10.54499/2022.07471.CEECIND/CP1718/
CT0001). TAM. acknowledges financial support from FCT under the CEEC Individual 2017, reference CEECIND/04530/2017/CP1393/CT0004 (DOI identifier: 10.54499/CEECIND/04530/2017/CP1393/CT0004), and from IT–Coimbra through the corresponding Assistant Research contract CT/Nº 004/2019-F00069.
Work by IS was supported by Grant Nº PID2021-129035NB-I00 funded
by MCIN/AEI/10.13039/501100011033 and by ERDF/EU.

\end{acknowledgments}

\appendix
\onecolumngrid

\section{Electro-optic permittivity} \label{sec:appendix_electro_optic}
According to Eqs. \eqref{E:conductivity_EO} and \eqref{E:permittivity_EO}, the EO susceptibility  is explicitly given by:
\begin{align}  
\db{\eps}_\text{EO} 
&=\frac{1}{-i \omega}\left( -\boldsymbol{\omega}_0\cdot   \db{D} \times \vec{1}_{3\times3}
+ \frac{1+i\omega\tau}{1+(\omega\tau)^2}\boldsymbol{\omega}_0   \times \db{D}^T \right)
\end{align}
Taking into account that $\boldsymbol{\omega}_0$ is a real vector, that $ \boldsymbol{\omega}_0\cdot   \db{D} \times \vec{1}_{3\times3}$ is anti-symmetric, and that 
$\left(\boldsymbol{\omega}_0   \times \db{D}^T\right)^T=-\db{D}\times\boldsymbol{\omega}_0$, we find that
\begin{align}  
\db{\eps}_\text{EO}^\dagger &=\frac{1}{i \omega}\left(\boldsymbol{\omega}_0\cdot   \db{D} \times \vec{1}_{3\times3}
- \frac{1-i\omega\tau}{1+(\omega\tau)^2}    \db{D} \times\boldsymbol{\omega}_0\right)
\end{align}
It then follows that the Hermitian and anti-Hermitian parts of the susceptibility are, respectively:
\begin{align}  
\db{\eps}_\text{EO}'&=\frac{\db{\eps}_\text{EO}   + \db{\eps}_\text{EO}^\dag }{2} 
=\frac{i}{2 \omega}\left(-2 \boldsymbol{\omega}_0\cdot   \db{D} \times \vec{1}_{3\times3} 
+ \frac{1+i\omega\tau}{1+(\omega\tau)^2}\boldsymbol{\omega}_0   \times \db{D}^T   
+ \frac{1-i\omega\tau}{1+(\omega\tau)^2}    \db{D} \times\boldsymbol{\omega}_0\right) \label{E:epsPrime}\\
%
%
\db{\eps}_\text{EO}''&=\frac{\db{\eps}_\text{EO}   - \db{\eps}_\text{EO}^\dag }{2i}
= \frac{1}{2 \omega}\left(  \frac{1+i\omega\tau}{1+(\omega\tau)^2}\boldsymbol{\omega}_0   \times \db{D}^T  
- \frac{1-i\omega\tau}{1+(\omega\tau)^2}    \db{D} \times\boldsymbol{\omega}_0\right)  \label{E:epsSecond}
\end{align}
The real part of  \eqref{E:epsPrime} 
gives Eq. \eqref{E:epsPrimeReal} of the main text. The imaginary part of  \eqref{E:epsPrime} is
\begin{align}  
\db{\eps}_\text{EO,I}'
&=\frac{1}{2 \omega}\left(-2 \boldsymbol{\omega}_0\cdot   \db{D} \times \vec{1}_{3\times3} 
+ \frac{1 }{1+(\omega\tau)^2}\left[ \boldsymbol{\omega}_0   \times \db{D}^T   
+ \db{D} \times\boldsymbol{\omega}_0\right] \right)\\
&=\frac{1}{2 \omega}\left(-2 \boldsymbol{\omega}_0\cdot   \db{D} \times \vec{1}_{3\times3} 
- \frac{1 }{1+(\omega\tau)^2} \left[\db{D}^T \cdot \boldsymbol{\omega}_0 \right] \times \vec{1}_{3\times3}  \right)
 \end{align}
where we used the identity $\boldsymbol{\omega}_0   \times \db{D}^T +\db{D} \times \boldsymbol{\omega}_0 = \left( -\db{D}^T \cdot \boldsymbol{\omega}_0 + {\rm{Tr}}\left\{ {\db{D}} \right\} \boldsymbol{\omega}_0 
\right) \times \boldsymbol{1}_{3\times3} $ and took into account that ${\db{D}}$ is traceless. Then, using that
$\left[\db{D}^T \cdot \boldsymbol{\omega}_0 \right] \times \vec{1}_{3\times3}= \boldsymbol{\omega}_0 \cdot \db{D}  \times \vec{1}_{3\times3}$ the latter expression simplifies to \eqref{E:epsPrimeImag} of the main text.
A similar decomposition of the real and imaginary parts of \eqref{E:epsSecond} leads to equations \eqref{E:epsSecondReal} and \eqref{E:epsSecondImag} of the main text.

\section{Berry curvature dipole tensors \label{Dsymmetry}}
 
Here, we present the explicit forms of the BD tensor for the crystallographic point groups that possess a principal axis. The tensor forms are given in the conventional Cartesian setting of the BCS. The tensor for the missing groups ($1$ and $m$) can be obtained with the BCS TENSOR program \cite{Gallego2019}. The tensor $\db{D}$, being a non-magnetic pseudotensor, shares the same structure as the magnetotoroidic tensor and can be expressed in the BCS (using Jahn’s notation) as eV$^2$.   \\

\begin{tabular}{l >{\centering\arraybackslash}p{15cm}}
Category A: & $\db{D}_{2}=
\begin{pmatrix}
D_{xx} &D_{xy}&0\\
D_{yx}&D_{yy}  & 0 \\
0&0 &-(D_{xx}+D_{yy})
\end{pmatrix}, \quad  \db{D}_3=\db{D}_4=\db{D}_6=
\begin{pmatrix}
D_{xx} &D_{xy}&0\\
-D_{xy}&D_{xx}  & 0 \\
0 &0 &-2D_{xx} 
\end{pmatrix}$. \\[0.7cm]
Category B: 
&$\db{D}_{mm2}=
\begin{pmatrix}
0 &D_{xy}&0\\
D_{yx}&0 & 0 \\
0 &0 &0
\end{pmatrix}.$ \\[0.7cm]
Category C: & $\db{D}_{3m}=\db{D}_{4mm}=\db{D}_{6mm}=
\begin{pmatrix}
0 &D_{xy}&0\\
-D_{xy}&0  & 0 \\
0 &0 &0
\end{pmatrix}$. \\[0.7cm]
Category D: & $\db{D}_{222}=
\begin{pmatrix}
D_{xx} &0&0\\
0&D_{yy} & 0 \\
0 &0&-(D_{xx}+D_{yy})
\end{pmatrix}, \quad \db{D}_{32}=\db{D}_{422}=\db{D}_{622}=
\begin{pmatrix}
D_{xx} &0&0\\
0&D_{xx}  & 0 \\
0 &0 &-2D_{xx} 
\end{pmatrix}$. \\[0.7cm]
Category E: & $\db{D}_{\overline{4}}=
\begin{pmatrix}
D_{xx} &D_{xy}&0\\
D_{xy}&-D_{xx}  & 0 \\
0 &0 &0
\end{pmatrix}, \quad \db{D}_{\overline{4}2m}=
\begin{pmatrix}
D_{xx} &0&0\\
0&-D_{xx}  & 0 \\
0 &0 &0
\end{pmatrix}$.
\end{tabular}

%

\section{Transfer matrix for propagation along $z$\label{AppendR}}

Here we derive the transfer (ABCD) matrix that characterizes wave propagation in the considered birefringent non-Hermitian materials.

The Maxwell's equations in the frequency domain are
 \begin{subequations}\label{E:Maxwell_equations}
\begin{align}
\nabla\times \vec{E}&= i\omega \mu_0 \vec{H} \\
\nabla\times \vec{H}&= -i\omega  \eps_0 \db{\eps} \cdot \vec{E}
\end{align}
 \end{subequations}
For propagation along the $z$-direction, we have $\nabla= \partial/\partial z \,\hat{\vec{z}}$, and these equations can be written in a matrix form as
\begin{align} \label{E:Maxwell_Schrodinger_equation}
i \frac{\partial}{\partial z}  \vec{f}(z)=\db{\vec{M}} \cdot \vec{f}(z)
\end{align}
where the four-component state vector  $\vec{f}$ is defined by
\begin{align}
\vec{f}(z)= \begin{pmatrix}E_x(z) & E_y(z) & H_x(z) & H_y(z) \end{pmatrix}^T
\end{align}
with $E_x$, $E_y$, $H_x$, $H_y$ the $x$ and $y$ components of the electric and magnetic fields at position $z$. The $4\times4$ matrix $\db{\vec{M}}$ is given by
\begin{align}
\db{\vec{M}}=\begin{pmatrix}
0 & 0 &  0 & -\omega \mu_0\\
0 & 0 & \omega \mu_0 & 0\\
 \omega  \eps_0  \eps_{yx} & \omega  \eps_0  \eps_{yy}  & 0  & 0\\
-\omega  \eps_0 \eps_{xx} & -\omega  \eps_0 \eps_{xy}  & 0 & 0
           \end{pmatrix}.
\end{align}
It is implicit that the $z$-direction is a symmetry axis of the system so that $\varepsilon_{xz}=\varepsilon_{yz}=\varepsilon_{zx}=\varepsilon_{zy}=0$.
Similar to a Schr\"odinger equation, the solution of Eq. \eqref{E:Maxwell_Schrodinger_equation}  is
\begin{align}\label{E:solutions_f}
\vec{f}(z)= e^{-i z \db{\vec{M}} }\cdot \vec{f}(0),
\end{align}
where the exponential $e^{-iz\db{\vec{M}}}$ is a $4\times4$ matrix that plays the role of a transfer matrix. The matrix exponential can be evaluated analytically (not shown here), or numerically. Thus, the fields at a generic position $z$ within the material depend only on the fields calculated at $z=0$.

\section{Reflection matrix for normal incidence\label{AppendRr}}

Here we derive the reflection matrix $\db{\vec{R}}$ for a generic dielectric material slab of thickness $d$ terminated by a perfect electric conducting (PEC) wall. It is assumed that a plane wave propagates along the $+z$-direction and impinges on the slab along the normal direction.
The region $z<0$ is filled by an isotropic dielectric with relative permittivity $\varepsilon_d$, whereas the region $0<z<d$ is occupied by the material slab.
It is implicit that the $z$-direction is a symmetry axis of the material.
%
%

To derive the expression for $\db{\vec{R}}$, we start by noting that the fields at the PEC wall ($z=d$) are necessarily of the form:
\begin{align}
\vec{f}(d)&= H_x(d)\hat{\vec{u}}_3+ H_y(d) \hat{\vec{u}}_4
\end{align}
where $H_x(d)$ and $H_y(d)$ are the components of the magnetic field at $z=d$. Using the transfer matrix [Eq. \eqref{E:solutions_f}],  the fields at $z=0$ can be written as $\vec{f}(0)=e^{i d \db{\vec{M}} }\cdot \vec{f}(d)$, or equivalently:
\begin{align}
\begin{pmatrix}E_x \\E_y \\H_x \\H_y \end{pmatrix}_{z=0}=\begin{pmatrix}
 \hat{\vec{u}}_1 \cdot e^{i d \db{\vec{M}} }\cdot \hat{\vec{u}}_3  & & \hat{\vec{u}}_1 \cdot e^{i d \db{\vec{M}} }\cdot \hat{\vec{u}}_4 \\ \hat{\vec{u}}_2 \cdot e^{i d \db{\vec{M}} }\cdot \hat{\vec{u}}_3  & & \hat{\vec{u}}_2 \cdot e^{i d \db{\vec{M}} }\cdot \hat{\vec{u}}_4 \\ \hat{\vec{u}}_3 \cdot e^{i d \db{\vec{M}} }\cdot \hat{\vec{u}}_3  & &\hat{\vec{u}}_3 \cdot e^{i d \db{\vec{M}} }\cdot \hat{\vec{u}}_4 \\\hat{\vec{u}}_4 \cdot e^{i d \db{\vec{M}} }\cdot \hat{\vec{u}}_3  & & \hat{\vec{u}}_4 \cdot e^{i d \db{\vec{M}} }\cdot \hat{\vec{u}}_4 \end{pmatrix} \cdot \begin{pmatrix}
 H_x(d) \\ H_y(d) \end{pmatrix} \label{E:f_zero_kildal}
\end{align}
where $\hat{\vec{u}}_i=(\delta_{i1}\;\delta_{i2}\;\delta_{i3}\;\delta_{i4})^T$ with $\delta_{ij}$ the Kronecker delta symbol, is a vector with the entry "1" at the $i$-th line.
Following the procedure outlined in Appendix A of Ref. [\onlinecite{lannebere_chiral_2025}], which uses an impedance matrix $\db{\vec{Z}}$ that relates the transverse components of the electric field $\vec{E}_t$ and the magnetic field $\vec{H}_t$ at the interface ($z=0^+$) as $ \db{\vec{J}}  \cdot \vec{E}_t= \db{\vec{Z}} \cdot \vec{H}_t$ \cite{Morgado2016,silveirinha_fluctuation_2018,Latioui2019}, it can be shown that the reflection matrix is given by:
\begin{align} \label{E:reflection_matrix}
 \db{\vec{R}} &= \left( \frac{\sqrt{\eps_d}}{\eta_0}  \db{\vec{J}}  \cdot \db{\vec{Z}}  \cdot \db{\vec{J}} + \db{\vec{1}}_{2\times 2}\right)^{-1} \cdot \left( \frac{\sqrt{\eps_d}}{\eta_0}  \db{\vec{J}}  \cdot \db{\vec{Z}}  \cdot \db{\vec{J}} - \db{\vec{1}}_{2\times 2} \right)
\end{align}
where $\eta_0$ is the free-space impedance, $\vec{1}_{2 \times 2}$ is the $2 \times 2$ identity matrix, $\db{\vec{J}}=\begin{pmatrix} 0 & 1 \\-1 & 0 \end{pmatrix}$ and 
\begin{align} \label{E:Z_matrix}
\db{\vec{Z}} &=\db{\vec{J}} \cdot \begin{pmatrix}
 \hat{\vec{u}}_1 \cdot e^{i d \db{\vec{M}} }\cdot \hat{\vec{u}}_3  & \hat{\vec{u}}_1 \cdot e^{i d \db{\vec{M}} }\cdot \hat{\vec{u}}_4 \\ \hat{\vec{u}}_2 \cdot e^{i d \db{\vec{M}} }\cdot \hat{\vec{u}}_3  & \hat{\vec{u}}_2 \cdot e^{i d \db{\vec{M}} }\cdot \hat{\vec{u}}_4 \end{pmatrix} \cdot  \begin{pmatrix} \hat{\vec{u}}_3 \cdot e^{i d \db{\vec{M}} }\cdot \hat{\vec{u}}_3  & \hat{\vec{u}}_3 \cdot e^{i d \db{\vec{M}} }\cdot \hat{\vec{u}}_4 \\\hat{\vec{u}}_4 \cdot e^{i d \db{\vec{M}} }\cdot \hat{\vec{u}}_3  & \hat{\vec{u}}_4 \cdot e^{i d \db{\vec{M}} }\cdot \hat{\vec{u}}_4 \end{pmatrix}^{-1}.
\end{align}

\twocolumngrid

\bibliography{Biblio_final}
\end{document}